\documentclass[]{aastex631}
\usepackage{color}
\usepackage{graphicx}
\usepackage{amsmath}
\usepackage{amssymb}
\usepackage[super]{nth}
\usepackage[ruled,vlined]{algorithm2e}

\newcommand{\M}{{\cal M}}

\shorttitle{DIMC}
\shortauthors{Steinberg \& Heizler}
\begin{document}

\title [mode = title]{A New Discrete Implicit Monte Carlo Scheme for Simulating Radiative Transfer Problems}
\author[0000-0003-0053-0696]{Elad Steinberg}
\email{elad.steinberg@mail.huji.ac.il}
\affiliation{Racah Institute of Physics, The Hebrew University, 9190401 Jerusalem, Israel}
\author[0000-0002-9334-5993]{Shay I. Heizler}
\affiliation{Racah Institute of Physics, The Hebrew University, 9190401 Jerusalem, Israel}
\email{shay.heizler@mail.huji.ac.il}

\begin{abstract}
We present a new algorithm for radiative transfer—based on a statistical Monte Carlo approach—that does not suffer from teleportation effects, on the one hand, and yields smooth results, on the other hand. Implicit Monte Carlo (IMC) techniques for modeling radiative transfer have existed from the 1970s. When they are used for optically thick problems, however, the basic algorithm suffers from "teleportation" errors, where the photons propagate faster than the exact physical behavior, due to the absorption-blackbody emission processes. One possible solution is to use semianalog Monte Carlo, in its new implicit form (ISMC), which uses two kinds of particles, photons and discrete material particles. This algorithm yields excellent teleportation-free results, but it also produces noisier solutions (relative to classic IMC), due to its discrete nature. Here, we derive a new Monte Carlo algorithm, Discrete Implicit Monte Carlo (DIMC), which also uses the idea of two kinds of discrete particles, and thus does not suffer from teleportation errors. DIMC implements the IMC discretization and creates new radiation photons for each time step, unlike ISMC. Using the continuous absorption technique, DIMC yields smooth results like classic IMC. One of the main elements of the algorithm is the avoidance of the explosion of the particle population, by using particle merging. We test the new algorithm on 1D and 2D cylindrical problems, and show that it yields smooth, teleportation-free results. We finish by demonstrating the power of the new algorithm on a classic radiative hydrodynamic problem—an opaque radiative shock wave. This demonstrates the power of the new algorithm for astrophysical scenarios.

\end{abstract}

\section{Introduction}
%Write introduction, talk about IMC and ISMC, maybe also talk about drawing photons from last absorption site scheme.

Radiation transfer is an integral part of many astrophysical processes, especially when it is coupled to hydrodynamic schemes. For example, the process of star formation from giant molecular clouds is heavily dependent on radiation pressure~\citep[e.g.]{star1, star2}. Accretion disks surrounding compact objects~\citep[e.g.]{disk}, whose dynamical properties depend on radiation matter coupling. Calculation of the emerging light-curve from supernovae (SN) explosions~\citep[e.g.]{sn} and shock breakout from SN~\citep[e.g.]{break}. It is also a major part in modeling high energy density physics and inertial confinement fusion~\citep{lindl}, thus often the benchmarks of radiative transfer are common in both fields~\citep{stone_bench}.

The basic equation that describes the motion of the photons and their interaction with the matter is the Boltzmann equation, or the transport equation for photons (also called the radiative transfer equation - RTE), coupled to the material energy equation~\citep{pomraning1973}. In the case of radiative hydrodynamics problems, it is a part of the energy-balance equation~\citep{zeldovich}. The simplest approximation for radiative transfer, is the widely-used diffusion approximation (also known as the Eddington approximation), which is much more trivial to solve than the full Boltzmann equation, especially in multi-dimensions. However, this approximation is valid when the angular distribution of the photons is isotropic (or close to isotropic), which tends to be valid in optically-thick media, close to local thermodynamics equilibrium (LTE). In optically thin media, the isotropic assumption may break, and high-order approximation are needed to model the exact problem. Flux-limiters (FL) or variable Eddington factors (VEF) may partially solve this issue~\citep{levpom,olson99}, however, these solutions are not universal, and depend on the specific problem that is solved (different FLs or VEFs fitted for different problems).

One of the well-known methods for solving radiative transfer problems is using statistical Monte-Carlo techniques, specifically the implicit Monte-Carlo (IMC) of~\citet{IMC}. This implicit algorithm enables the use of large time-steps, replacing the process of absorption and black-body emission in opaque materials, with an {\em{effective}} isotropic scattering term. This algorithm, with some modifications and improvements is being used during the last five decades~\citep{four_decades,Gentile2001,Densmore2011,teleportation,teleportation2,Gentile2010,Gentile2011,Gentile2012,Gentile2014,MCCLARREN_URBATSCH,shi_2020_1,shi_2020_3,shi_2019,gael2}. However, the basic algorithm suffers from a well-known numerical error, that causes teleportation in optically-thick materials~\citep{teleportation,teleportation2,ISMC}. When  photons enter an opaque cell (specifically, a ``cold" cell), they are absorbed in the outer part of the numerical cell and heat the material. Afterwards, in the next time step, the heated cell emits photons uniformly (in the simplest choice) from the cell, which causes the photons to propagate faster than they should, i.e., teleport. Of course, this numerical issue converges with spatial resolution and time-step, however, in practice (especially in high dimensions), the convergence is slow~\citep{ISMC,Steinberg1}. Specifically we mention that using a fixed spatial resolution, $\Delta x$, \emph{decreasing} $\Delta t$ further and further gives rise to a wrong ``converged", teleported solution~\cite{Steinberg1}.

There are several algorithms that reduce teleportation-errors, that rely on reconstructing the spatial profile of the temperature inside the cells (``source tilting")~\citep{tilt, Wollaeger2016, Smedley-Stevenson}. Another alternative is a hybrid IMC-diffusion method~\citep{Densmore} where optically thick regions are modeled as diffusion regions. Recently, \cite{teleportation} devised a method to create photons from the locations of absorption sites in the previous time step, if known. All of these methods reduce the teleportation error, but none of them manage to eliminate it altogether.

One particular promising method to eliminate the teleportation error, named Semi-Analog MC (SMC), was introduced in~\cite{SMC}. In this method, there are two types of particles, radiation photons and material particles. Instead of creating radiation photons from the thermal energy of the cell, in this scheme material particles have a chance per unit time to transform into radiation photons, and when radiation photons are absorbed they are transformed into material particles. Unfortunately, the explicit nature of the scheme forced very small time steps and made the method unappealing~\citep{SMC_stable}. A breakthrough was introduced recently by the milestone work of~\cite{ISMC}, where an implicit version of SMC (ISMC) was derived. \citet{Steinberg1} have explored the ISMC algorithm in many one dimensional and two dimensional problems (both in XY and RZ symmetries), both in gray and multi-group benchmarks. The ISMC algorithm exhibited no teleportation errors compared to the IMC algorithm, and thus, converged much faster in spatial resolution and time-step.

Nevertheless, the ISMC results yield noisier solutions, due to the discrete nature of the absorption-emission process~\citep{ISMC,Steinberg1}. Although it achieves a fast convergence in spatial resolution and time-step, ISMC requires many more particles for yielding IMC-like smooth results. In addition, in cases where there is a large ratio between the heat capacities of the radiation and the material, ISMC suffers from low statistic errors, and may require a prohibitively large number of particles.

In this work we present a new algorithm, Discrete Implicit Monte-Carlo (DIMC), that shares the advantages of both schemes: It has no teleportation errors in finite spatial resolution and time-step due to the use of the basic idea of SMC, using two kinds of particles (photons and material particles). In addition, it yields smooth results as IMC (while achieving comparable run time as IMC), using the continuous absorption algorithm~\citep{IMC,milagro,four_decades}, and classic IMC implicitization. Smooth results are important even more, when the radiative transfer module is coupled to hydrodynamics module, as in astrophysical problems.
The nature of DIMC requires an algorithm that prevents the population of particles (especially material particles that are created continuously from the continuous absorption of radiation photons) to `explode'.

This work is structured as follows: In Sec.~\ref{imc_ismc} we present a brief review about the IMC and the ISMC implicit algorithms. In Sec.~\ref{dimc} we present in detail the new DIMC algorithm, and its sub-modules. In sec.~\ref{results} we present some well-known pure radiative transfer benchmarks, both 1D and 2D, presenting smooth teleportation-free results. Next, in Sec.~\ref{shocks} we present the problem of radiative shocks, where the radiative transfer is coupled to hydrodynamics, in the well-known problem of radiative shocks~\citep{shock,gentile2019,nick_mmc}. We show that in optically-thick problems, the new algorithm presents excellent results also in full radiative hydrodynamics problems, compared to the other algorithms.

\section{Review of IMC and ISMC algorithms}
\label{imc_ismc}
In their seminal work,~\cite{IMC} have laid the foundations for implicit Monte-Carlo(IMC), that is the de facto MC method for the past 50 years. 
The basic equations of radiation transfer are the Boltzmann equation for photons, coupled to an energy-balance material energy equation~\citep{IMC}:
\begin{subequations}
\label{eq:org_rad_transfer}
\begin{eqnarray}
    \frac{1}{c}\frac{\partial I(\boldsymbol{r},\boldsymbol{\Omega},t)}{\partial t}+\boldsymbol{\Omega\cdot\nabla} I(\boldsymbol{r},\boldsymbol{\Omega},t)&=&-\left(\sigma_a+\sigma_s\right)I(\boldsymbol{r},\boldsymbol{\Omega},t)+\sigma_a B(T)+ \sigma_s\int_{4\pi}\frac{I(\boldsymbol{r},\boldsymbol{\Omega'},t)}{4\pi}\boldsymbol{d\Omega'}\\
    \frac{\partial e(T)}{\partial t}&=&c \sigma_a\left(\int_{4\pi}I(\boldsymbol{r},\boldsymbol{\Omega'},t)\boldsymbol{d\Omega'}-4\pi B(T)\right)
    \end{eqnarray}
\end{subequations}
where $I(\boldsymbol{r},\boldsymbol{\Omega},t)$ is the radiation intensity for a unit volume and unit direction , $\boldsymbol{\Omega}$, at time $t$, $\sigma_a$ is the absorption cross-section (opacity), $\sigma_s$ is the scattering cross-section (which in this work, we assume is isotropic, such as Thomson scattering), $e(T)$ is the thermal energy per unit volume of the medium, $T$ is the material temperature, $B(T)=aT^4/4\pi$ is the frequency integrated Planck function, $a$ is the radiation constant and $c$ is the speed of light. 
The basic idea behind IMC, is to define the Fleck parameter,
\begin{equation}
    f=\frac{1}{1+\beta \sigma_a c  \Delta t}
\end{equation}
where $\beta\equiv\frac{\partial B}{\partial e}$ is the ratio between the radiation and material heat capacities~\citep{IMC} and $\Delta t$ is the discretized time-step. By finite differencing Eq.~\ref{eq:org_rad_transfer}(b), one can rewrite Eqs. \ref{eq:org_rad_transfer} as: 
\begin{subequations}
\label{eq:rad_transfer}
\begin{eqnarray}
    \frac{1}{c}\frac{\partial I(\boldsymbol{r},\boldsymbol{\Omega},t)}{\partial t}+\boldsymbol{\Omega\cdot\nabla} I(\boldsymbol{r},\boldsymbol{\Omega},t)&=&-\left(\sigma_a+\sigma_s\right)I(\boldsymbol{r},\boldsymbol{\Omega},t)+f\sigma_a B(T)+\left(\left(1-f\right)\sigma_a +\sigma_s\right)\int_{4\pi}\frac{I(\boldsymbol{r},\boldsymbol{\Omega'},t)}{4\pi}\boldsymbol{d\Omega'}\\
    \frac{\partial e(T)}{\partial t}&=&fc \sigma_a\left(\int_{4\pi}I(\boldsymbol{r},\boldsymbol{\Omega},t)\boldsymbol{d\Omega}-4\pi B(T)\right)
    \end{eqnarray}
\end{subequations}

The introduction of the Fleck parameter, creates a new ``effective" scattering term that arises from photon absorption and re-emission within the time step. The algorithm, is called implicit Monte-Carlo (IMC) since it uses implicit value of $B$ inside the Boltzmann equation, and thus enables using practical (large-enough) time steps, even when optically-thick materials are involved. This implicitization of the radiation transfer equation has been very successfully and is widely used~\citep{four_decades,Gentile2001,Gentile2010,Gentile2011,Gentile2012,Gentile2014,MCCLARREN_URBATSCH,shi_2020_1,shi_2020_3,shi_2019,gael2}. As mentioned in the introduction, IMC unfortunately suffers from what is known as the ``teleportation" error. When the optical depth is large and/or large optical depth gradients occur, insufficient spatial resolution can lead to an un-physical solution, where the heat wave travels too fast. This error arises from the fact that in IMC photons are created uniformly within a cell, even when physically the heat wave has propagated only a fraction of the cell (usually the outer part of a cold cell)\citep{MCKINLEY, four_decades}. The magnitude of the error also depends on the size of the time step, where time steps that are \emph{smaller increase} the magnitude of the error. In a given spatial (finite) resolution, this issue may lead to a convergence to a wrong answer~\citep{ISMC,Steinberg1}.

The semi-analog Monte Carlo (SMC) algorithm~\cite{SMC}, uses the idea of two kind of particles, photons and material particles. The photons propagate with speed of light and can be absorbed or scattered, while the material particles have zero-velocity, and can ``propagate" only by flipping to a photon-particle, due to black-body radiation emission. This means that a photon can be created only where material particles are present, and since the material particles are created only where a photon was absorbed, teleportation is, by definition, avoided. The SMC was not a very practical method, until the milestone work of~\citet{ISMC}, which offered a new {\em{implicit}} version, called implicit semi-analog Monte-Carlo (ISMC). The new implicitization is based on defining $\eta(e)\equiv 4\pi  B(e)/e$, and $\zeta\equiv\beta-\eta$ then the radiative transfer equation can rewritten as:
\begin{subequations}
\begin{eqnarray}
    \frac{1}{c}\frac{\partial I(\boldsymbol{r},\boldsymbol{\Omega},t)}{\partial t}+\boldsymbol{\Omega\cdot\nabla} I(\boldsymbol{r},\boldsymbol{\Omega},t)+(\sigma_a+\sigma_s)I(\boldsymbol{r},\boldsymbol{\Omega},t)&=&\frac{\sigma_a\eta\chi e}{4\pi}+\left(\sigma_s+\left(1-\chi\right)\sigma_a\right)\int_{4\pi}\frac{I(\boldsymbol{r},\boldsymbol{\Omega'},t)}{4\pi}\boldsymbol{d\Omega'}\\
    \frac{\partial e(T)}{\partial t}&=&c \sigma_a\chi\left(\int_{4\pi}I(\boldsymbol{r},\boldsymbol{\Omega},t)\boldsymbol{d\Omega}-\eta  e\right),
    \end{eqnarray}\label{eq:ismc}
\end{subequations}
with $(1-\chi)\sigma_a$ as an ISMC equivalent to the ``effective'' scattering term, that results from the implicit MC scheme, and $\chi\equiv\frac{1}{1+\zeta c\sigma_{a}\Delta t}$, in analogy to the Fleck factor in IMC (with $\zeta $ that replaces $\beta $ in IMC).

As the IMC algorithm, the new implicitization enables the use of large enough time-steps and reasonable simulation times, when optically-thick regions are involved in the problem. The ISMC algorithm was developed and tested in several problems, showing no teleportation errors, both in spatial resolution and time step in various one and two dimensional problems~\citep{ISMC,Steinberg1}. We note that on the other hand, the ISMC yields noisier material field temperature for a given statistics (number of particles), due to its discrete process of absorption and emission. We note, that it is possible to reduce the noise level in ISMC (as well as in IMC) using track length estimators, which average the physical quantities over the time step. Additionally, since there is a one to one correspondence between the transformation of a material particle to a radiation particle and vice versa, a large heat capacity ratio between the radiation and the material can lead to noisy results.

\section{Discrete Implicit Monte Carlo (DIMC)}
\label{dimc}
The driving rationale behind our new proposed method, Discrete Implicit Monte Carlo (DIMC), is to combine the teleportation free advantage of ISMC~\citep{ISMC} with the relatively smooth material field temperature of IMC~\citep{IMC}. We recognize that the representation of the material field with particles, allows ISMC to have an effectively ``infinite" spatial resolution for the position of the heat wave front, and prevents teleportation errors. On the other hand, the continuous energy deposition in IMC allows the material temperature field to be relatively smooth, even with a modest number of photons. Our aim, is to develop a method that on the one hand has a particle based representation of the material field, while on the other hand has continuous energy deposition. We would also like to decouple the energy scales between the radiation field and material field, so problems with large heat capacity ratio can be easily solved.

Our basic starting point are the gray IMC implicitization, Eq. \ref{eq:rad_transfer} (and not the ISMC implicitization, Eqs.~\ref{eq:ismc}). Thus, in DIMC, the thermal energy is a cell based quantity, however it is also represented by material particles, which are stationary and only contain location and energy. The sum of the energy of the material particles in a cell is always equal to the cell's thermal energy. During the initialization phase of DIMC, in addition to the standard initialization of IMC (i.e. creating photons and assigning the cell properties), we create a modest amount ($\sim 10-30$) of material particles uniformly in each cell, whose energy is equal to the cell's energy over the number of material particles in the cell.
We note that we have tried also the ISMC linearization and it gives very similar results. we have decided to stick with the IMC linearization since it is more studied and would be easier to convert existing IMC codes to our new algorithm.

As in IMC, at the beginning of each time step, new photons are created. While in IMC, new photons are created uniformly throughout the cell, in DIMC new photons are created only at locations of the material particles. Whenever a new photon is created, its energy is deducted from both the cell quantity and the material particle. An outline of the photon creation process is given in algorithm \ref{alg:create}.
\begin{algorithm}[t]
\label{alg:create}
\SetAlgoLined
 \ForAll{\emph{cells}}
 {
  Nphoton = \emph{Number of photons to create in cell}\;
  V = Volume of cell\;
  Ephoton = $f\sigma_a c B(T)\Delta t$ V / Nphoton\;
  MaterialParticlesInCell = \emph{An array of all the material particles in the cell}\;
  \For{$i=1$ \KwTo \emph{Nphoton}}
  {
    photon = new Photon()\;
    idx = \emph{Draw a random index from \emph{MaterialParticlesInCell} weighted by \emph{MaterialParticlesInCell.energy}}\;
    photon.location = MaterialParticlesInCell[idx].location\;
    photon.energy = Ephoton\;
    MaterialParticlesInCell[idx].energy $-=$ Ephoton\;
    \emph{Subtract} Ephoton \emph{from cell's energy}\;
    // Rest of photon properties (i.e. angle, time) are determined as in IMC\ 
  }
 }
 \caption{Photon Creation}
\end{algorithm}

Another modification with regard to IMC is the photon propagation. We are interested in a method to propagate photons that resembles IMC in the sense that it allows for continuous energy deposition~\citep{IMC,milagro,four_decades}, and also allows to effectively advance in time the material particle field. Every time a photon loses energy due to absorption, a history of that loss is recorded, and at the end of the time step or when a photon moves to a new cell, we create a new material particle, whose energy is the deposited energy and its location is drawn from an exponential distribution based on the effective absorption cross section. The full photon propagation algorithm is shown in algorithm \ref{alg:propogate}. For both our IMC and DIMC implementations, we present the values at the end of the time step and not a track length estimated one.

\begin{algorithm}[t]
\label{alg:propogate}
\SetAlgoLined
x = photon.location\;
EnergyForNewMaterialParticle = 0\;
 \While{\emph{photon}.\emph{time} $<$ \emph{time}  $+\Delta t$}
 {
 $r$ =  Random[0, 1]\;
 $d_{\mathrm{collision}}$ = $-\frac{\log{(1-r)}}{(1-f)\sigma_a+\sigma_s}$\;
 $d_{\mathrm{mesh}}$ = \emph{Distance for photon to reach cell boundary}\;
 $d_{\mathrm{time}}$ = \emph{speed of light} $\cdot$ (time + $\Delta t$ - photon.time )\;
 $d =$ min($d_{\mathrm{collision}}$, $d_{\mathrm{mesh}}$, $d_{\mathrm{time}}$)\;
    $\Delta E$ = photon.energy$\left(1-\exp{\left(-f\sigma_a d\right)}\right)$\;
    $r$ =  Random[0, 1]\;
    ExpRandom = $-\log{\left(1+r\left(\exp{\left(-f\sigma_a d\right)}-1\right)\right)} / f\sigma_a d$\;
    $x$\textsubscript{create} = photon.location + ExpRandom$\cdot\mathbf{\Omega}\cdot d$\;
    photon.location $+= \mathbf{\Omega}\cdot d$\;
    $x =$ $\left(\text{EnergyForNewMaterialParticle}\cdot x + \Delta E \cdot x\text{\textsubscript{create}}\right)$ / $\left(\text{EnergyForNewMaterialParticle} + \Delta E\right)$\;
   EnergyForNewMaterialParticle += $\Delta E$\;
   photon.energy $-=$ $\Delta E$\;
 \eIf{$d_{\mathrm{collision}}=d$}
 {
    // Do as in IMC (i.e. update photon time, draw new angle, update cell energy)
  }
  {
   Create new material particle with energy EnergyForNewMaterialParticle and location x\;
        EnergyForNewMaterialParticle = 0\;
    // Rest as in IMC (i.e. update photon time, update cell energy)
 }
 }
 \caption{Single Photon Propagation}
 \end{algorithm}

Due to the fact that each photon creates a new material particle each cycle, the amount of material particles must be limited is some manner. If not, the population of material particles and photons will explode quickly. Thus, a careful algorithm, that limits the total number of material particles per cell has to be applied. The chosen algorithm limits at the end of each time step, the total number of material particles in a cell to some pre-defined number, N\textsubscript{keep}, whose value is typically in the range 10-30. In each cell, we randomly select N\textsubscript{keep} of the material particles weighted by their energy, and then iterate over the remaining material particles in the cell. For each material particle we find the spatially closest particle from the particles that where randomly selected, and add to it the energy of the other particle while changing its location in a manner to conserve the pair's ``center of mass" (where here mass is the energy). The detailed algorithm is presented in algorithm \ref{alg:merge}.

\begin{algorithm}[t]
\label{alg:merge}
\SetAlgoLined
 \ForAll{cells}
 {
 MaterialParticlesInCell = \emph{An array of all the material particles in the cell}\;
  ParticlesToKeep = \emph{An array of length} N\textsubscript{keep} \emph{of material particles randomly selected weighted by their energy from} MaterialParticlesInCell\;
  \ForAll{\emph{m} = Material particle in \emph{MaterialParticlesInCell} but not in \emph{ParticlesToKeep}}
  {
    idx = \emph{Index of nearest particle in} ParticlesToKeep\;
    ParticlesToKeep[idx].location = $\left(\text{ParticlesToKeep}[\text{idx}].\text{location}\cdot \text{ParticlesToKeep}[\text{idx}].\text{energy} + \right.$ \
    $\left. \text{m}.\text{location}\cdot \text{m}.\text{energy}\right)$ / $\left(\text{ParticlesToKeep}[\text{idx}].\text{energy} + \text{m}.\text{energy}\right)$\;
    ParticlesToKeep[idx].energy $+=$ m.energy\;
   delete m\;
  }
 }
 \caption{Material Particle Merging}
\end{algorithm}

\section{Pure Radiative Transfer Benchmarks}
\label{results}
In this section we present results of our new DIMC algorithm, compared to the classic IMC and ISMC algorithms. We choose three different benchmarks, one is 1D and two are 2D (RZ geometry) benchmarks that produce significant teleportation errors in the classic IMC implementations~\citep{Steinberg1}: 1D Marshak-wave~\citep{teleportation,ISMC}, RZ Cylindrical Hohlraum~\citep{MCCLARREN_URBATSCH}, and the Graziani's ``crooked pipe" problem~\citep{Gentile2001,teleportation}.

\subsection{1D Marshak Wave}
\cite{ISMC} presented a simple 1D Marshak wave test problem that demonstrates the ``teleportation" issue using IMC. In this problem, the domain, $x\in[0,4]$, is divided into evenly spaced cells, the left boundary of the domain is a black body with a normalized temperature of unity, the opacity is set to be $\sigma_a = 10\cdot T^{-3}$ and the heat capacity is constant $C_V=7.14a$. The initial material temperature is set to be $T(t=0)=0.01$ and we run the simulation until a time $ct=500$. We run this problem varying the time step at a fixed spatial resolution, as well as varying the spatial resolution at a fixed time step. For all the runs we create a total of $50\times$(Number of cells) photons per time step and limit the total number of photons to be $200\times$(Number of cells). For the DIMC runs, we limit the number of material particles to be 15 per cell. Our results are compared to a solution obtained using diffusion approximation solver, which yields the correct behavior in this optically thick regime~\citep{ISMC}. 

\begin{figure}
\centering
(a)\includegraphics*[width=4.75cm]{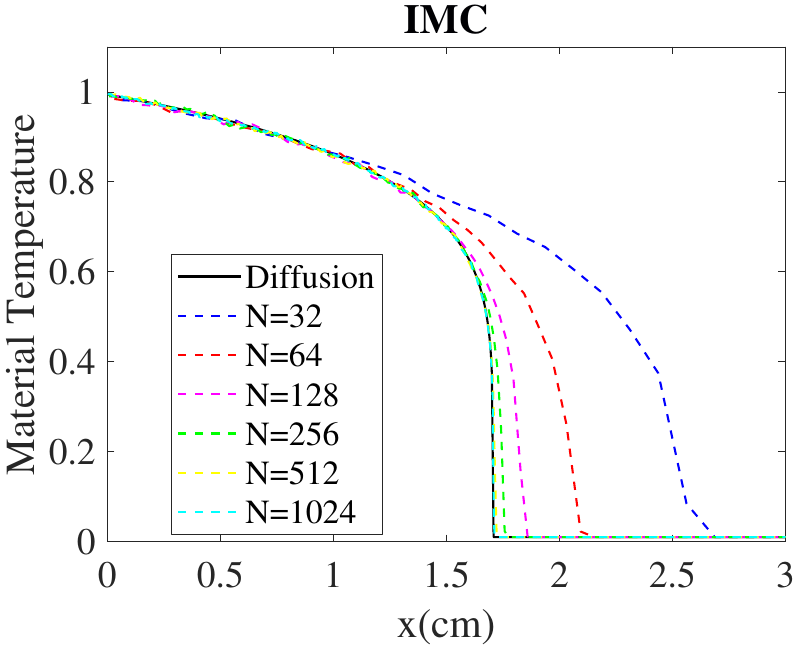}
(b)\includegraphics*[width=4.75cm]{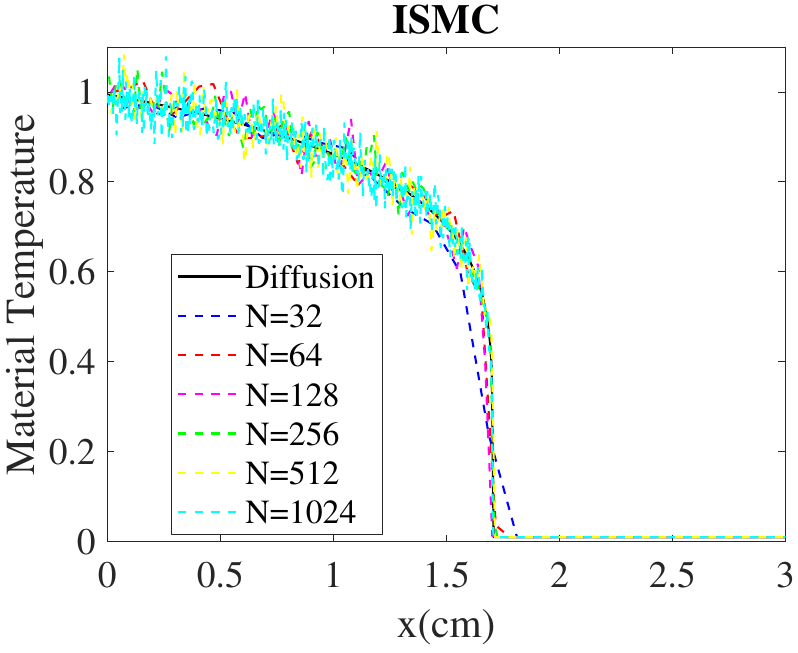}
(c)\includegraphics*[width=4.75cm]{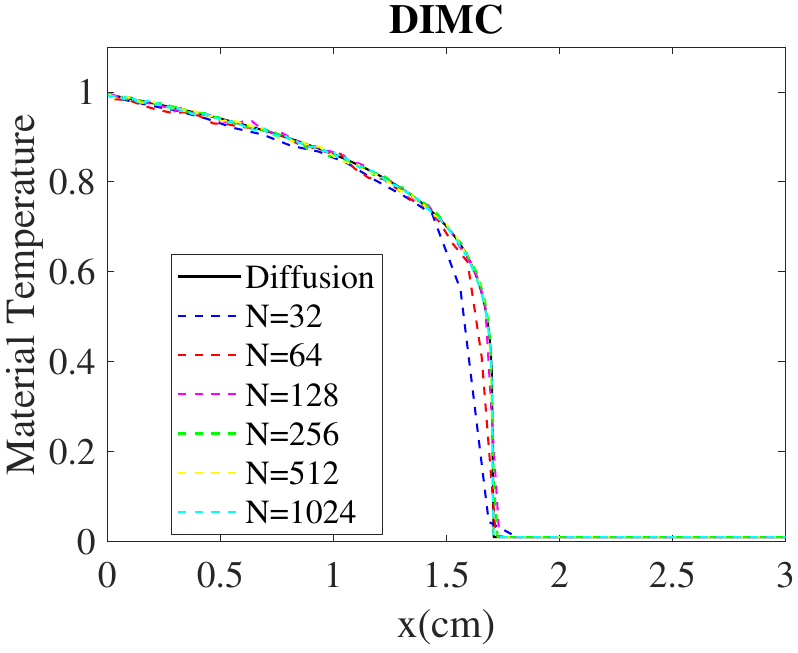}
\caption{The material temperature at time $ct=500$ for the Marshak wave test problem for various resolutions for the (a) IMC, (b) ISMC and (c) DIMC methods. All runs used a constant time step of $c\Delta t = 0.01$. The solution achieved using a diffusion solver is shown in black.}
\label{fig:marshak_res}
\end{figure}

Figure~\ref{fig:marshak_res} shows the material temperature for the different schemes using a fixed time step of $c\Delta t = 0.01$ while varying the spatial resolution. As expected from IMC, the front of the heat wave advances too fast if the spatial resolution is too low. %and/or the time step is too small.
On the other hand, ISMC does not suffer from any teleportation error, however, it has an increased noise level compared to the IMC runs. Our new method, DIMC, clearly retains the smoothness of the IMC results, while suffering from no teleportation error as in the ISMC runs.

\begin{figure}
\centering
(a)\includegraphics*[width=4.75cm]{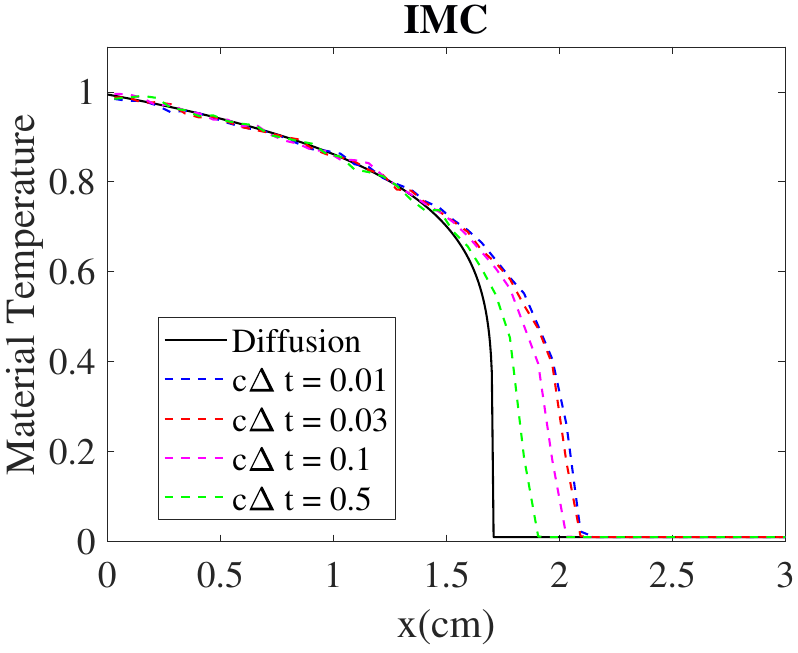}
(b)\includegraphics*[width=4.75cm]{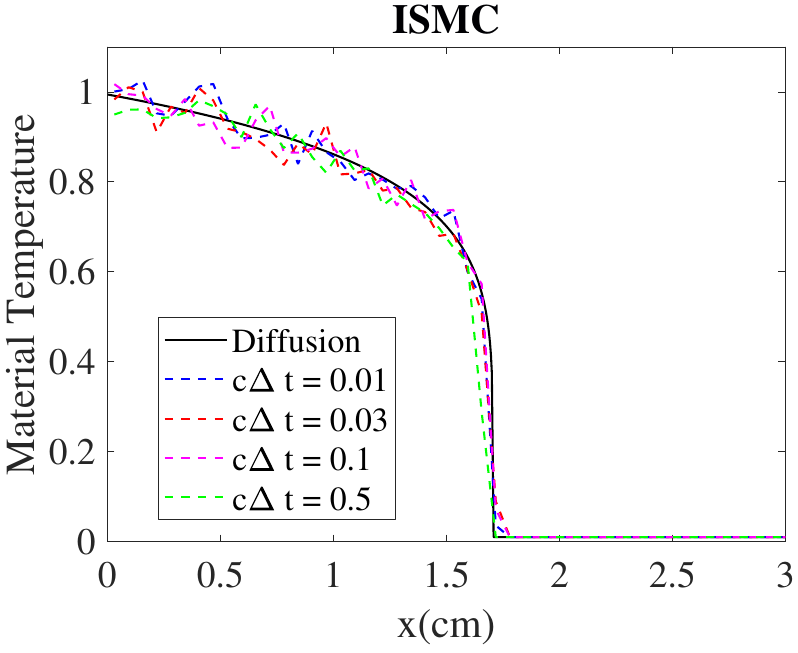}
(c)\includegraphics*[width=4.75cm]{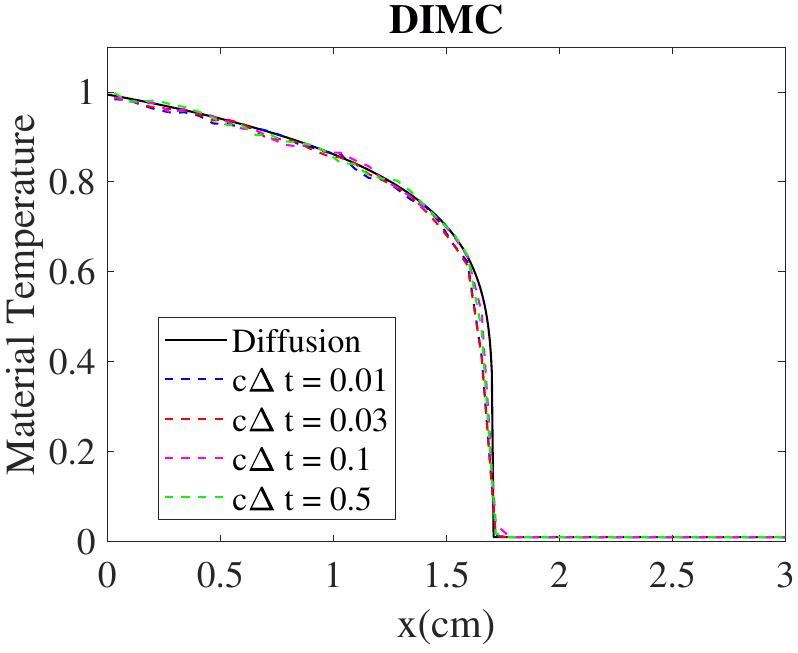}
\caption{The material temperature at time $ct=500$ for the Marshak wave test problem for various time steps for the (a) IMC, (b) ISMC and (c) DIMC methods. All runs used a uniform resolution of 64 cells. The solution achieved using a diffusion solver is shown in black.}
\label{fig:marshak_time}
\end{figure}

Moreover, for a fixed spatial resolution of 64 cells, we see in Fig.~\ref{fig:marshak_time} that both ISMC and DIMC give the same results without regard to the time step, while IMC suffers from an increased teleportation error as the time step is \emph{decreased}. Again, ISMC yields noisier results than DIMC (and also IMC). Compared to the IMC runs, the DIMC ran on average $2.5\%$ faster and the ISMC $30\%$ faster than the IMC runs (The reason that IMC is slower than DIMC is when a heat wave just starts to penetrate a cold cell, there are many effective scatterings due to the high opacity. In IMC most photons are created in the bulk and unergo many scatterings before losing their energy. In contrast, in DIMC, as long as the heat wave hasn't propagated significantly into the cell, most of the photons are created at the outer edge of the cold cell, and therefore have an increased chance to escape to a neighboring hot cell, where it will go less effective scatterings before being absorbed).

\subsection{McClarren \& Urbatsch (2009) 2D Cylindrical Hohlraum}
\begin{figure}
\centering
\includegraphics*[width=7.5cm]{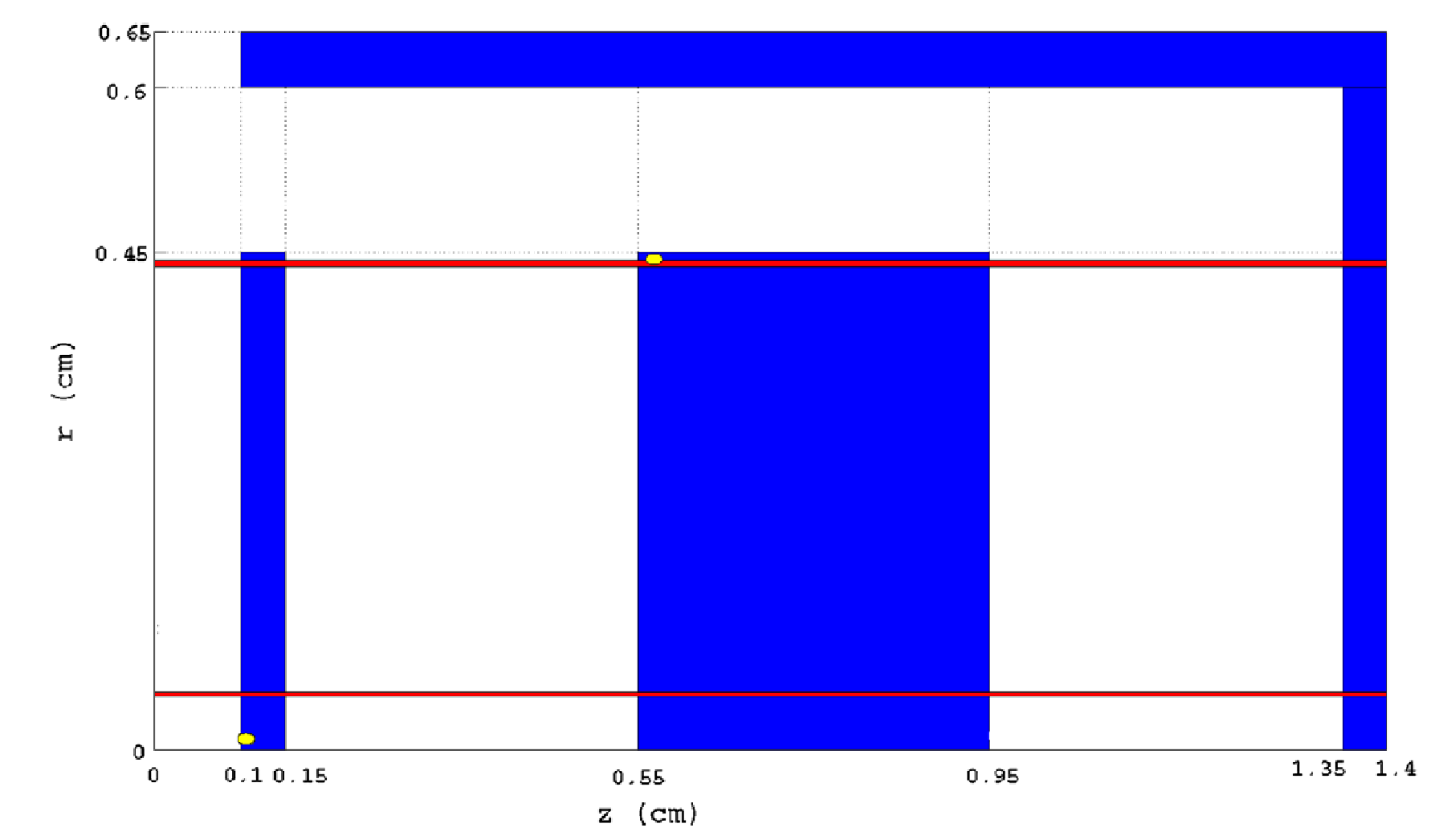}
\caption{The geometry of the~\citet{MCCLARREN_URBATSCH} cylindrical hohlraum problem. The absorbing material is depicted in blue and white is vacuum. The figure is taken from~\cite{MCCLARREN_URBATSCH}.}
\label{fig:RZ_geom}
\end{figure}

We proceed to a two-dimensional hohlraum-like benchmark problem. There are several hohlraum benchmarks, most of them, in XY geometry~\citep{brunner,MCCLARREN2,MCCLARREN_URBATSCH,shi_2020_1,shi_2020_3,shi_2019}. We are interested in the RZ geometry benchmark, because it suffers from teleportation significantly more than others~\cite{Steinberg1}, and thus, yields an excellent benchmark problem, to study the teleportation problem using different algorithms. As can be shown in Fig.~\ref{fig:RZ_geom}, the hohlraum benchmark problems in the literature are composed of an opaque outer walls that represents the hohlraum walls (made usually from gold) in an inner opaque ``capsule", that represents the ICF capsule in real HEDP modeling~\citep{lindl}. In almost all the benchmarks, the inner hohlraum volume is empty, (i.e., only contain only radiation energy and zero cross-sections).

\cite{MCCLARREN_URBATSCH} have presented a 2D cylindrical hohlraum problem, that has been investigated lately in several works \citep{shi_2020_1,shi_2020_3,shi_2019}. The hohlraum has a cylindrical radius of 0.65 cm and a length of 1.4 cm, the exact geometry and dimensions of the hohlraum are given in Fig.~\ref{fig:RZ_geom}. The blue represents the optically-thick absorbing material which has an opacity of $\sigma_a=300\left(T/\text{keV}\right)^{-3}\text{cm}^{-1}$, and a constant heat capacity of $C_V = 3\cdot 10^{15}\;\text{erg}/\text{keV}/\text{cm}^3$. %The white regions represents vacuum which has zero opacity and infinite heat capacity.
All of the boundaries are set to be vacuum except the $z=0$ boundary, where there is a black body source with a constant temperature of 1KeV. In order to show the strengths and weakness of the different methods, we vary both the spatial resolution and the time step. Our nominal run has a uniform mesh with a cell size of $\Delta r = \Delta z = 0.01\;\text{cm}$ and a constant time step of $\Delta t = 10^{-11}\;\text{s}$. In addition we perform a run with double the spatial resolution and a run with a small constant time step of $\Delta t = 10^{-12}\;\text{s}$. The system is evolved until a time of $t=10\;\text{ns}$ and the hohlraum is initially cold ($T=300$K). For all of the three methods we create $10^6$ new particles each time step, and limit the total number of particles to be $5\cdot10^6$. For the DIMC runs we limit the number of material particles to be 30 per cell.

\begin{figure}
\centering
(a)\includegraphics*[width=4.75cm]{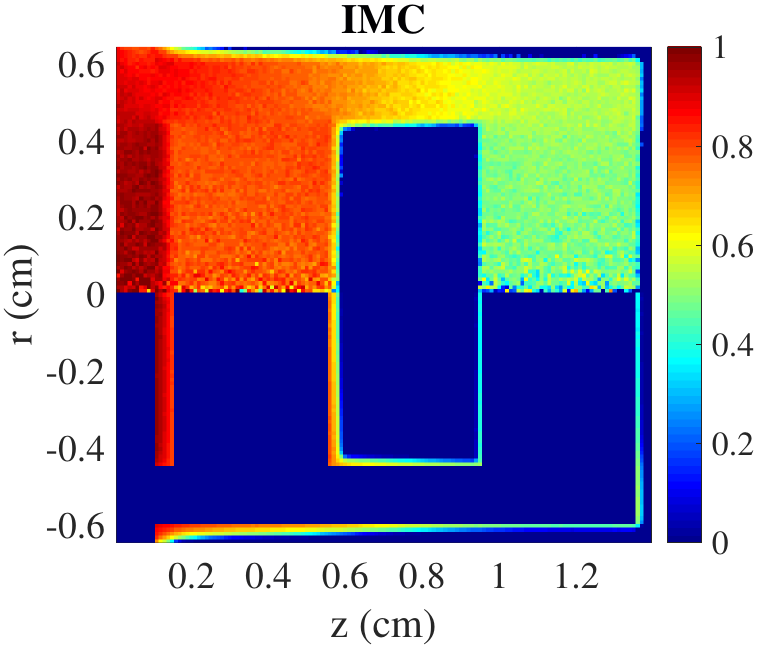}
(b)\includegraphics*[width=4.75cm]{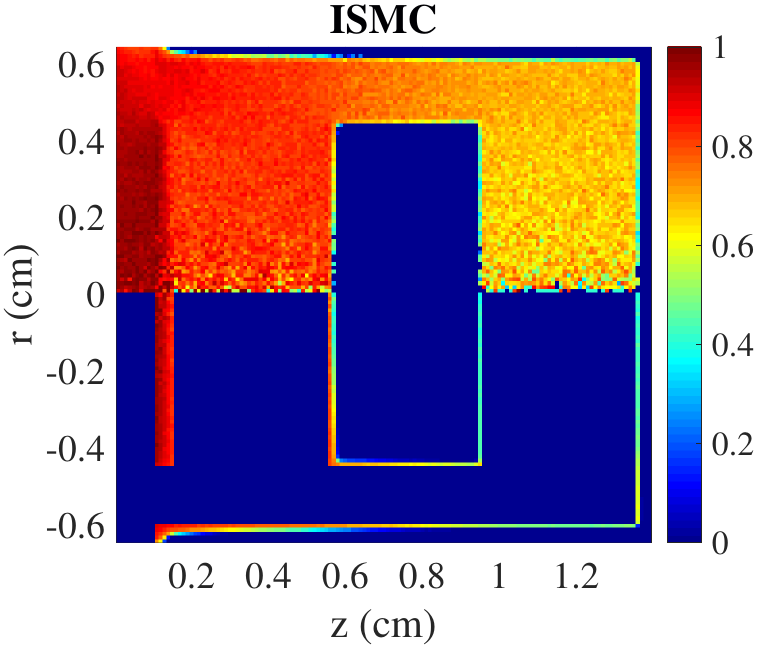}
(c)\includegraphics*[width=4.75cm]{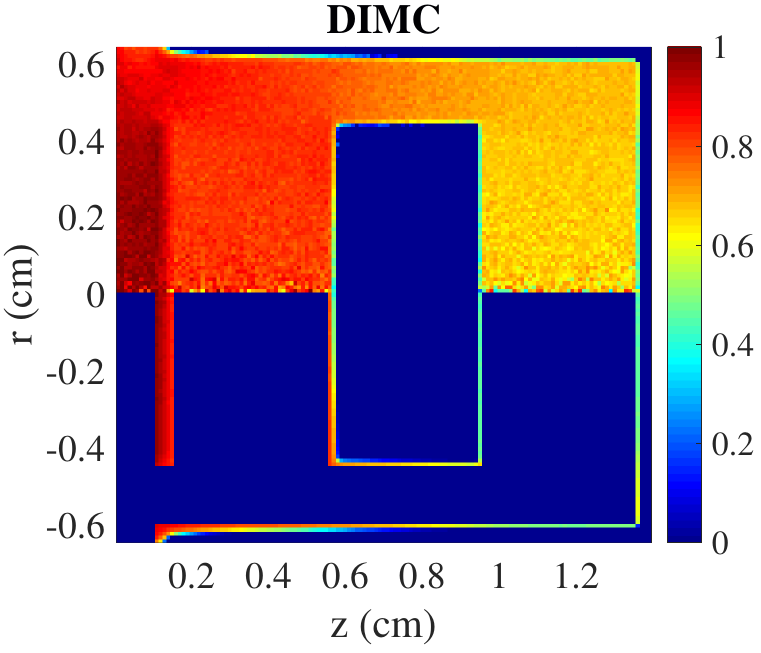}
\caption{The temperature map (in keV) for the cylindrical hohlraum test problem presented in \cite{MCCLARREN_URBATSCH} at time $t=1$ ns. The top half of the figure represents the radiation temperature and the bottom half represents the material temperature. The three different methods are represented in (a) IMC, (b) ISMC and (c) DIMC.}
\label{fig:cyl_holo_time}
\end{figure}
Fig.~\ref{fig:cyl_holo_time} shows the radiation (upper parts) and material (lower) temperatures for our nominal run at time $t=1$ ns for the three methods. The ISMC and DIMC methods get the correct temperature distribution while IMC has a cooler radiation temperature (upper part in Fig.~\ref{fig:cyl_holo_time}(a)). This is a direct result of the evident teleportation of heat into the opaque material that is seen in the IMC run (lower part in Fig.~\ref{fig:cyl_holo_time}(a)), and less energy is available for the hohlraum volume. Since in the ISMC run, a large fraction of the particles are in the material state (as opposed to radiation photons), the results yield a noisier radiation field compared to the DIMC run.

\begin{figure}
\centering
(a)\includegraphics*[width=6.5cm]{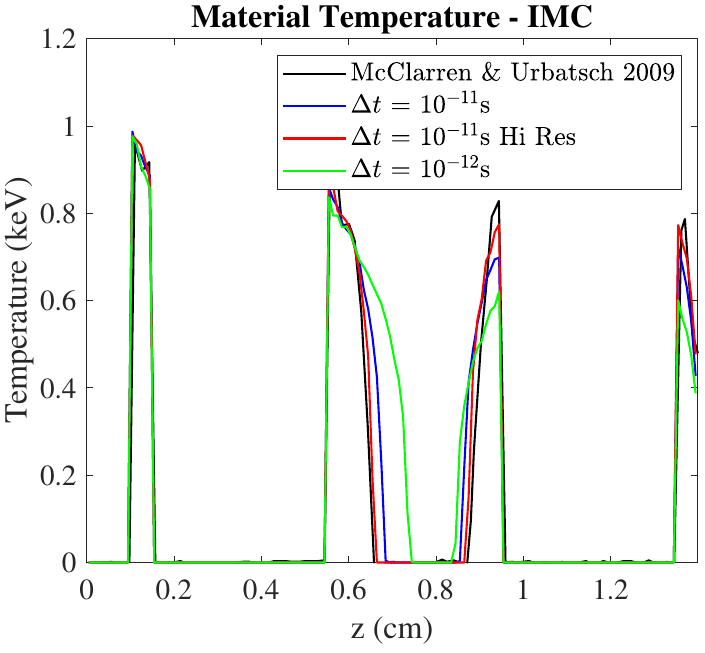}
(b)\includegraphics*[width=6.5cm]{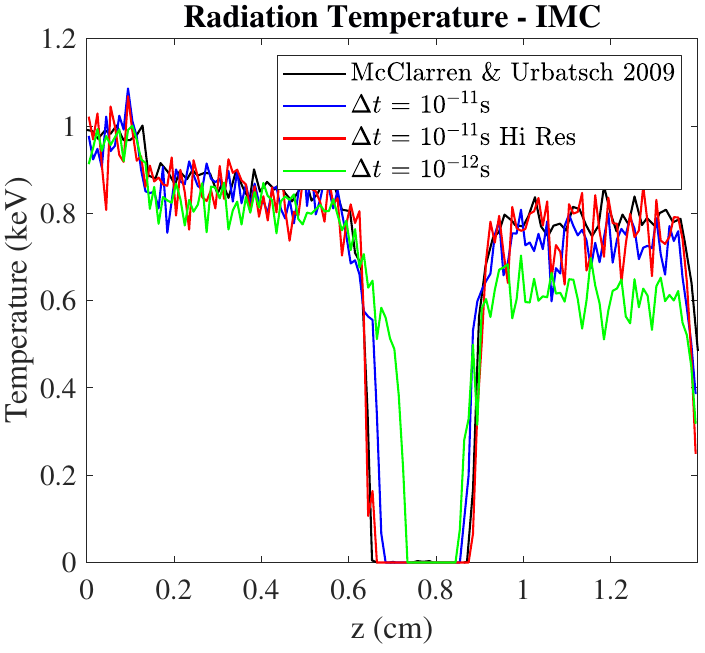}
(c)\includegraphics*[width=6.5cm]{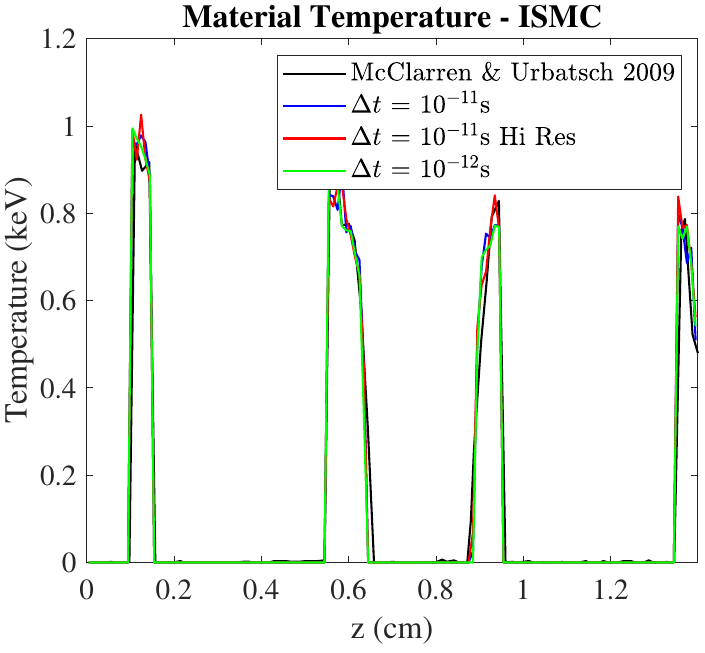}
(d)\includegraphics*[width=6.5cm]{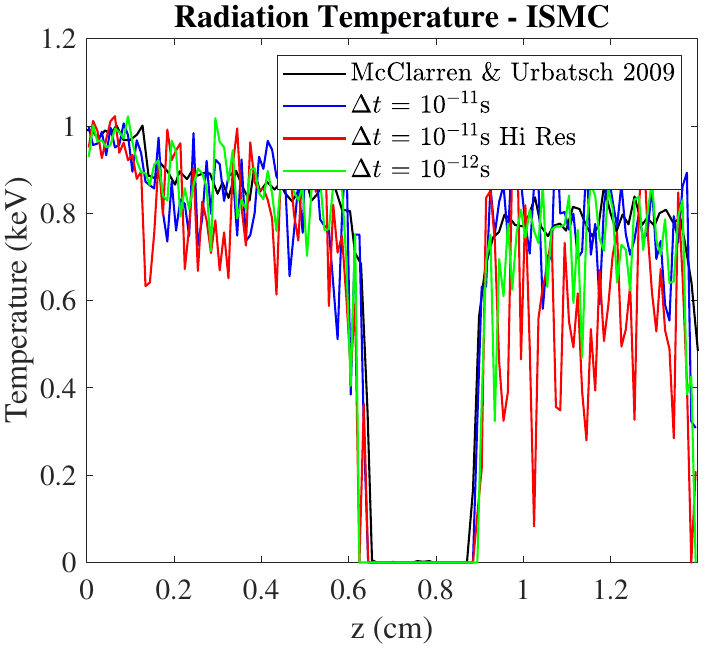}
(e)\includegraphics*[width=6.5cm]{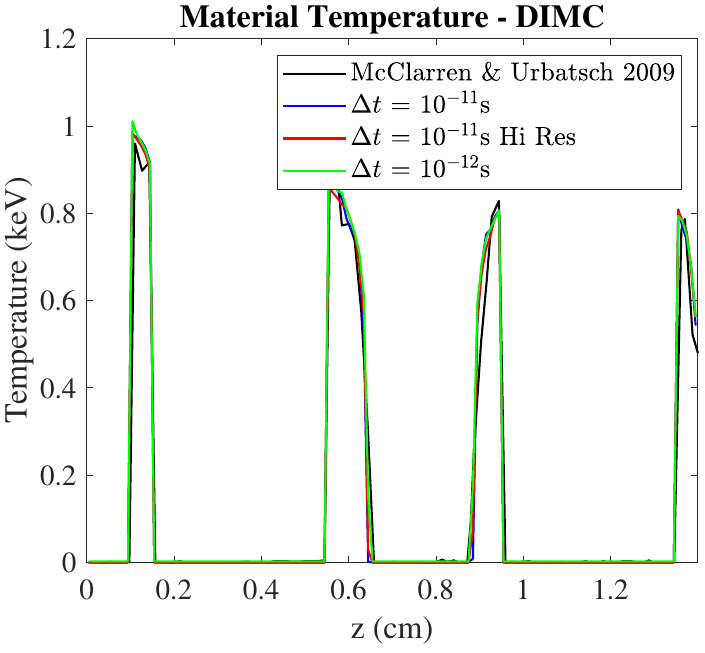}
(f)\includegraphics*[width=6.5cm]{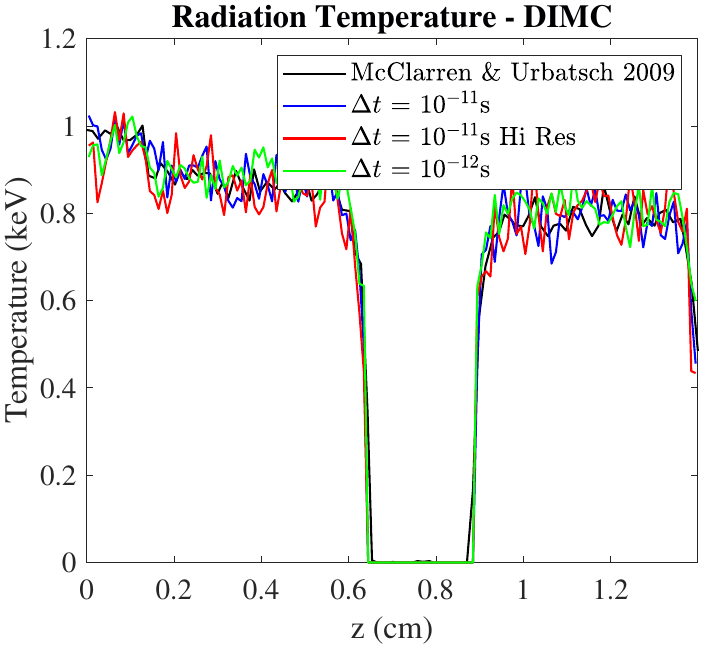}
\caption{The temperature along the $r=0.05\text{ cm}$ line at time $t=10{ ns}$ for the \cite{MCCLARREN_URBATSCH} cylindrical hohlraum problem using the IMC method. The material temperature is shown in panel (a) and the radiation temperature is shown in panel (b). (c+d) Same as Fig.~\ref{fig:cyl_holo_IMC} using the ISMC algorithm. (e+f) Same as Fig.~\ref{fig:cyl_holo_IMC} using the DIMC algorithm.}
\label{fig:cyl_holo_IMC}
\end{figure}
In Fig.~\ref{fig:cyl_holo_IMC}, we show the material and radiation temperature at time $t=10$ ns along the line $r=0.05$ cm for the three different methods. In the IMC case (Fig.~\ref{fig:cyl_holo_IMC}(a+b)), we see that decreasing the time-step causes significant teleportation and even the nominal run is not converged, while only the higher resolution run is converged. The ISMC method (Fig.~\ref{fig:cyl_holo_IMC}(c+d)) is converged for all the resolutions and time steps, but has an increased noise in the radiation field since a large fraction of the particles are in the material particles and not as radiation photons. Our new method, DIMC (Fig.~\ref{fig:cyl_holo_IMC}(e+f)), shows converged results and low noise for all of its runs.

\subsection{Graziani's ``crooked-pipe" benchmark}
\begin{figure}
\centering
\includegraphics*[width=10cm]{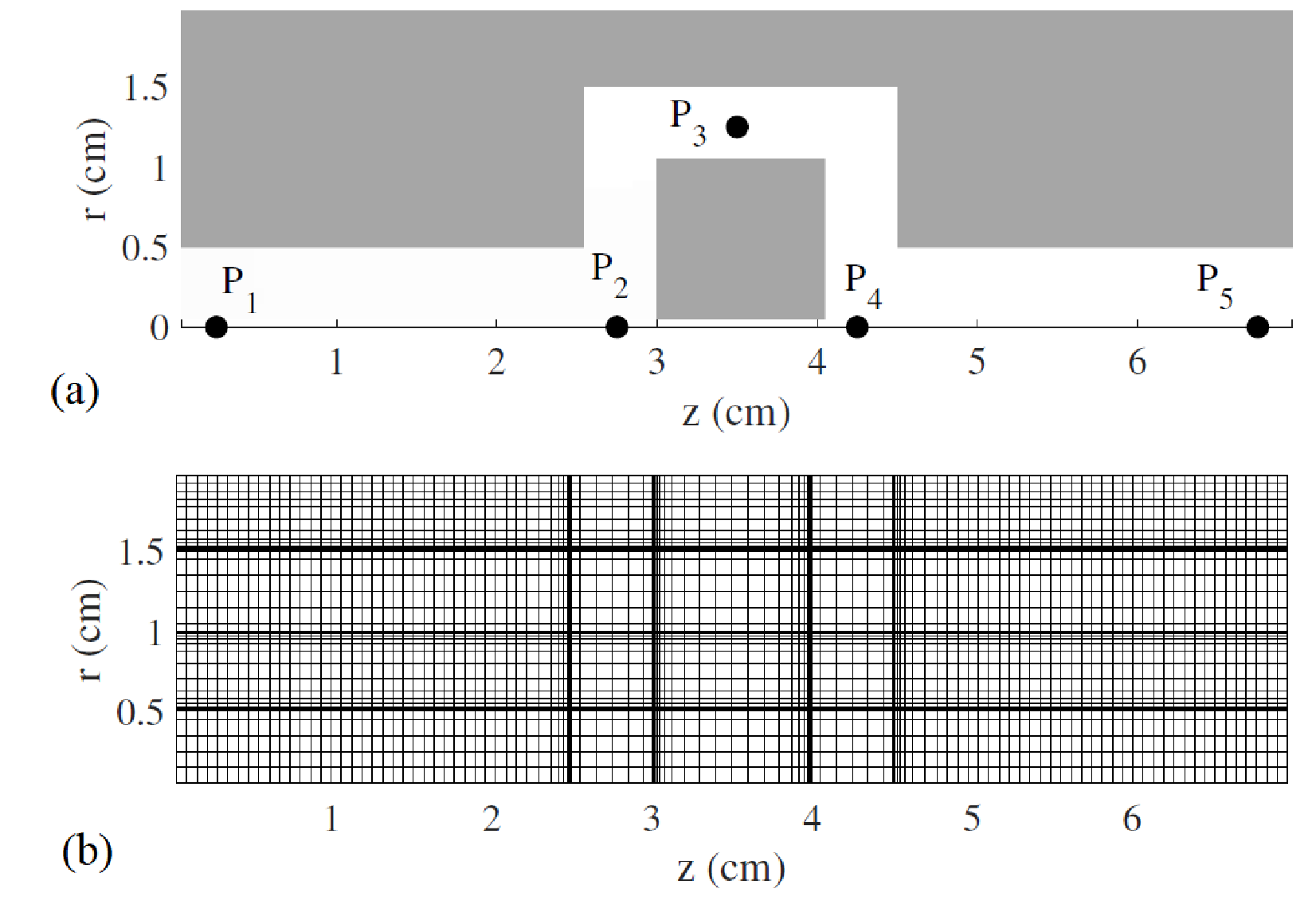}
\caption{(a) The geometry for the crooked pipe test problem. The high opacity material is in gray and the low opacity material is in white. (b) The grid used in the problem.}
\label{fig:pipe_geo}
\end{figure}

The Graziani's ``crooked-pipe" test~\citep{Gentile2001,teleportation} measures the ability of a radiative transfer code to follow the radiation along a complex radiation-flow geometry with a four-magnitude discontinuous opacities between the optically thin and thick zones. In this test, a black-body source with a temperature of 0.5KeV is located at the edge of the optically-thin material, that radiates toward the optically-thin cylindrical ``crooked" pipe. The exact geometry of the test is presented in Fig.~\ref{fig:pipe_geo}(a). This problem suffers from a severe teleportation errors, and thus creates a good opportunity to test the different algorithms, in a complex radiation-flow scenario.

The optically-thick material, depicted in gray, has a constant absorption opacity of $\sigma_a=2000\text{ cm}^{-1}$ and a constant heat capacity of $C_V = 10^{16}\;\text{erg}/(\text{keV cm}^3)$. The optically-thin material, depicted in white, has a constant absorption opacity of $\sigma_a=0.2\text{ cm}^{-1}$ and a constant heat capacity of $C_V = 10^{13}\;\text{erg}/(\text{keV cm}^3)$. Initially, all of the domain has a cold temperature of 0.05 KeV and the time step is set to be $10^{-11}\text{ s}$. The time step is then increased by a factor of 1.1 in each time step until it reaches $10^{-8}$s.
The spatial resolution is set to $\Delta r=\Delta z = 0.01$cm but due to the high opacity discontinuity, a 10 cell logarithmically spaced grid is added at the interfaces between the two materials, as shown in Fig.~\ref{fig:pipe_geo}(b). The size of the smallest cell in the logarithmically spaced zone (the first opaque cell) is set to be either $1.5\cdot 10^{-3}\text{ cm}$ (high spatial resolution) or $4\cdot 10^{-3}\text{ cm}$ (low spatial resolution).
For the IMC and DIMC methods we create a total of $10^6$ photons each time step and there is no need to limit the number of photons since the majority of them are absorbed within a single time step. On the other hand, the large ratio between the heat capacities of the material and the radiation in this problem, forced us to increase the number of new particles in the ISMC method to $10^7$ and limit the total number of particles to $2\cdot 10^8$ in order to get any reasonable statistics. We note that this is a major drawback of the ISMC method, where the ratio between particles in the photon state and the material state is set by the ratio of the heat capacities (as well as the fact that there is no continuous energy deposition by the photons in ISMC, as mentioned before). In the DIMC runs we use 30 material particles per cell.

\begin{figure}
\centering
(a)\includegraphics*[width=5.5cm]{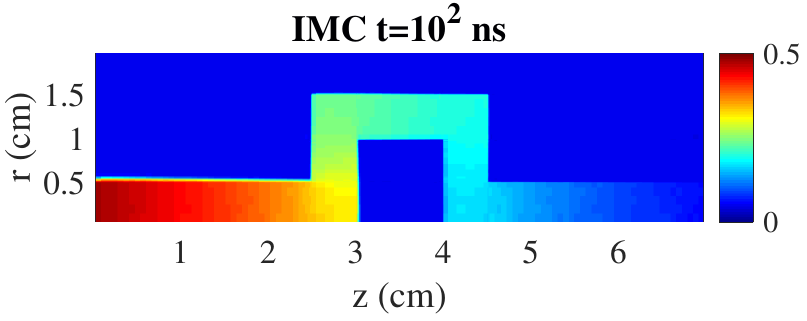}
(b)\includegraphics*[width=5.5cm]{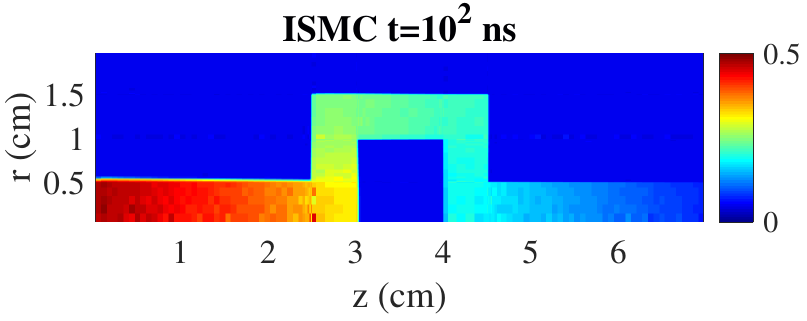}
(c)\includegraphics*[width=5.5cm]{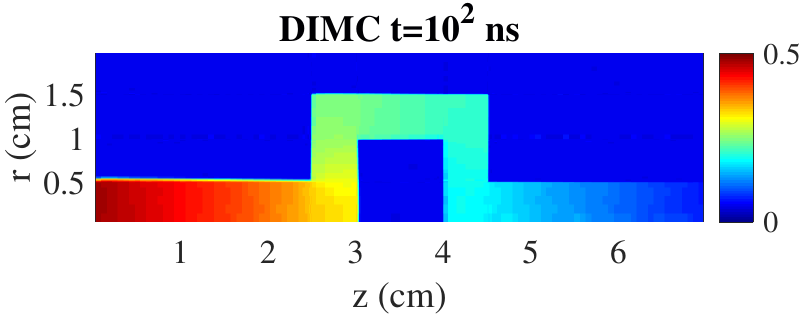}
(d)\includegraphics*[width=5.5cm]{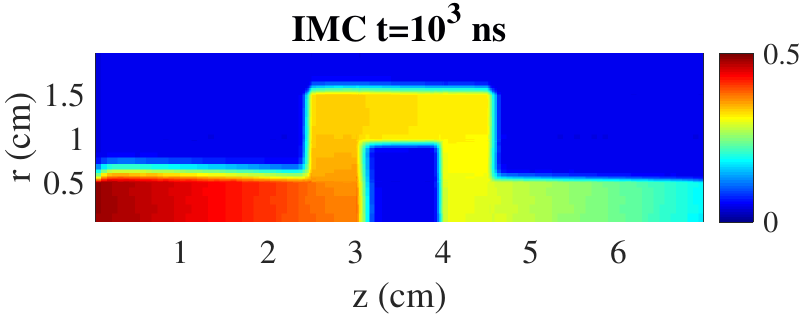}
(e)\includegraphics*[width=5.5cm]{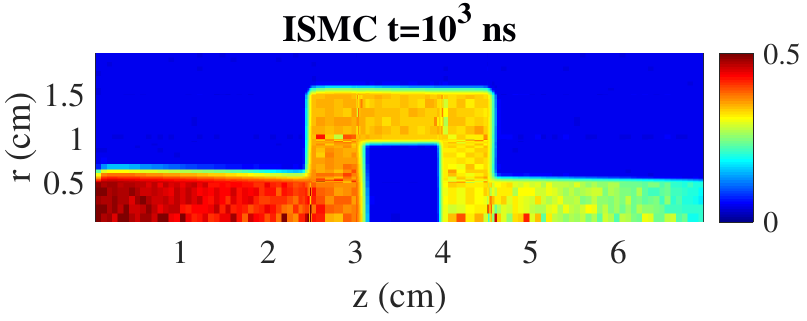}
(f)\includegraphics*[width=5.5cm]{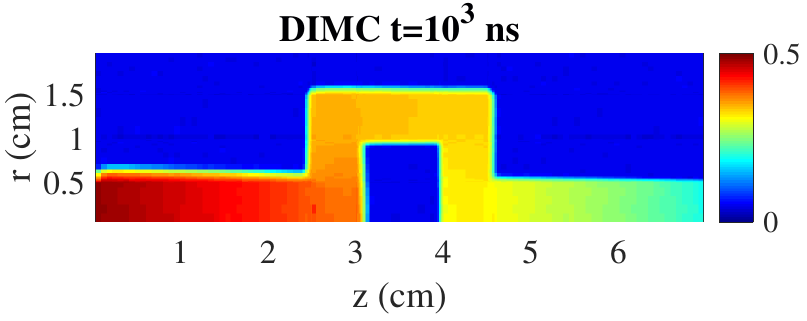}
\caption{The material temperature (keV) for different times materials and the three different methods. The figures show the results from the runs where the smallest cell size is $4\cdot 10^{-3}\text{ cm}$.}
\label{fig:pipe_colormap}
\end{figure}

In Fig.~\ref{fig:pipe_colormap} we present a temperature maps in two representative times, for the different schemes. We can see that ISMC yields noisier results than both IMC and DIMC schemes, where as the IMC scheme suffers from enhanced energy teleportation toward the opaque material. To quantify the results, we measure the material temperature as a function of time at five different points around the inner opaque material (see Fig.~\ref{fig:pipe_geo}(a)): $(r=0,\;z=0.25),\;(r=0,\;z=2.75),\;(r=1.25,\;z=3.5),\;(r=0,\;z=4.25)$ and $(r=0,\;z=6.75)$ 

\begin{figure}
\centering
(a)\includegraphics*[width=7.5cm]{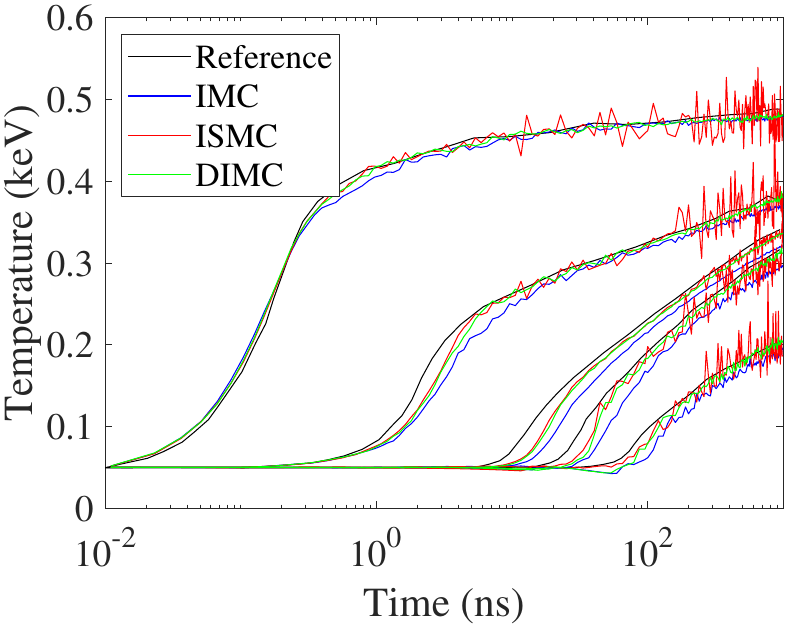}
(b)\includegraphics*[width=7.5cm]{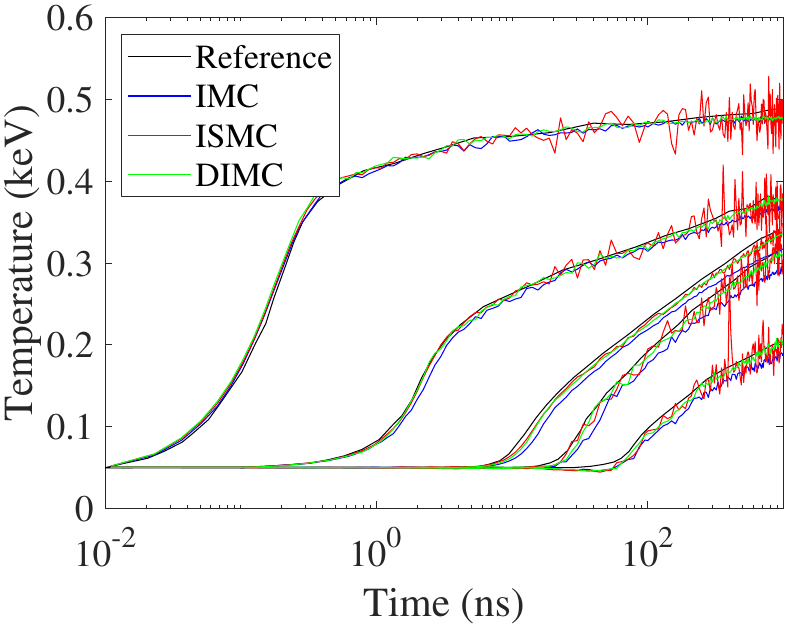}
(c)\includegraphics*[width=8cm]{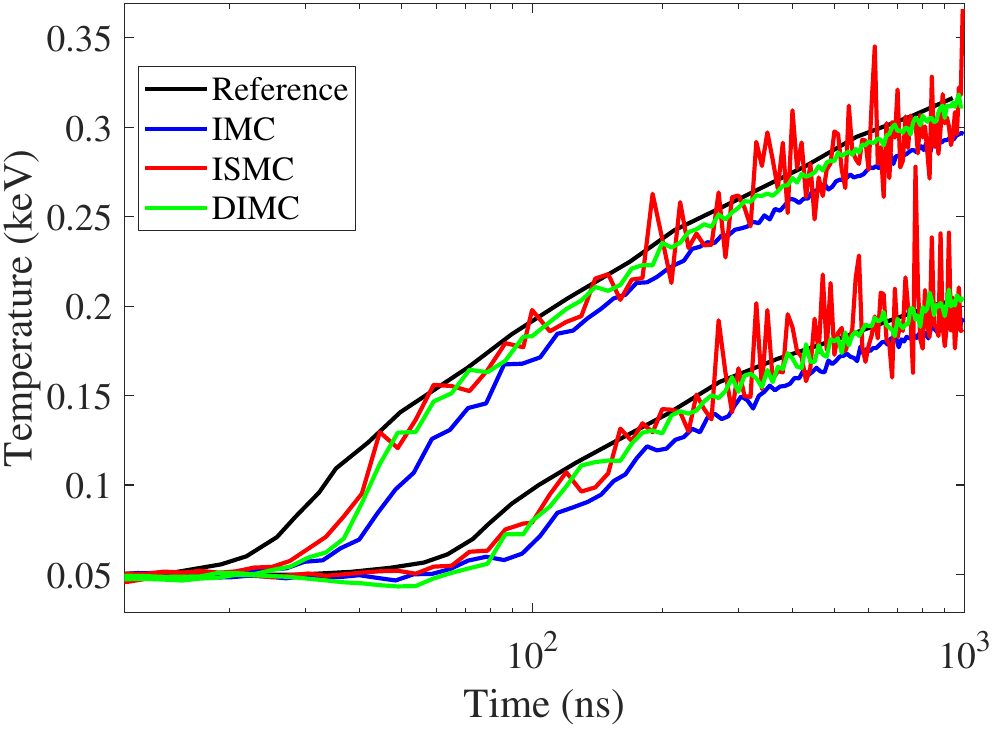}
\caption{The material temperature as a function of time for the five reference points, as described in the text. The results for IMC are shown in blue, ISMC in red, DIMC in green and the reference solution taken from \citet{Gentile2001} in black. The temperature is shown for grids in which the smallest cell in the boundary layer between the high opacity and low opacity materials, is (a) $4\cdot 10^{-3}\text{ cm}$ or (b) $1.5\cdot 10^{-3}\text{ cm}$. (c) A zoom in of the last two points from (a).}
\label{fig:pipe_result}
\end{figure}

Fig.~\ref{fig:pipe_result} shows the material temperature as a function of time for the five measured points for all three methods for the low spatial resolution (Fig.~\ref{fig:pipe_result}(a)) and high spatial resolution (Fig.~\ref{fig:pipe_result}(b)) runs. In the high resolution run, all three methods get the right over all result, but ISMC exhibits considerably more noise than IMC and DIMC. Inspecting the low resolution run, we see in Fig.~\ref{fig:pipe_result}(c) that a better convergence is achieved with DIMC compared to the IMC run, a result of less teleportation of radiation into the opaque material.

\section{Radiating Shock Waves benchmarks}
\label{shocks}
So far we have examined pure radiative transfer problems, showing that the DIMC is a teleportation-free algorithm, converges as fast as ISMC, while showing similar numerical noise as the classic IMC algorithm (which is far better than the numerical noise of ISMC). In this section we generalize the examination for a radiative-hydrodynamics problem, radiative shock waves in optically thick regions. Specifically, we choose a very relevant example for astrophysical application, which is the radiating shock wave problem of~\cite{shock}. \cite{shock} have calculated a semi-analytic solution for a steady-state radiating shock wave using a non-equilibrium (2-Temp) diffusion approximation. In sufficiently opaque problems, the diffusion approximation should yield a close enough result with the exact transport behavior.

We couple our MC implementations to a Lagrangian second order Godunov hydrodynamic scheme using a simple operator-split method as described in~\cite{nick_mmc,gentile2019}. Since we are interested in optically thick problems, we have implemented the minimal material motion corrections (MMC), ignoring aberrations and Doppler shift effects, i.e, we drag the photons with the fluid velocity, adjust the photons energy using the gradient of velocity term ($e_p^{n+1}=e_p^n[\textrm{Volume}^n/\text{Volume}^{n+1}]^{1/3}$) and add a radiation pressure term ($P_{\mathrm{Rad}}=U_{\mathrm{Rad}}/3$) to the momentum balance equation.
The main addition to both the ISMC and DIMC algorithms, is the coupling of the material particle propagation in the presence of hydrodynamic motion. After the hydro step is finished, the material particles within a cell are dragged along with the fluid's velocity. If the cell's internal energy has increased, we create a few new material particles with the added energy uniformly within in the cell, in order to maintain the equality between the cell's internal energy and the sum of the material particles energy within a cell. If the cell's internal energy has been reduced, we reduce a constant fraction from all of the material particles energy in order to maintain once again the equality of energies.

We use the fully detailed implementation of~\cite{gentile2019}, which introduce a Monte-Carlo solution for both Mach 2 ($\M=2$) problem, when the radiation and material temperatures differ significantly and non-LTE transport effects occur, and high Mach 45 problem ($\M=45$), where the radiation and material temperatures are almost in LTE.
First, we examine the correctness of the new DIMC, as well as our implementation for the IMC and ISMC schemes in these two benchmarks. Next, we finish by introducing an optically-thick $\M=45$-like problem, that demonstrates the teleportation of IMC, and its absence in the ISMC and DIMC implementations.

\subsection{Mach 2 ($\M=2$) Case}
We start by simulating the $\M=2$ shock, where the radiation and the gas are not in equilibrium throughout the test, and a significant presence of a true hydrodynamical shock can be seen. In this benchmark, the gas is assumed to be an ideal gas with an adiabatic index of $\gamma=5/3$ and a constant specific heat capacity of $1.91\cdot 10^8\textrm{ erg/g}\cdot\text{K}$. The absorption opacity is a function of the material temperature $\sigma_a=0.362\rho (T/\text{keV})^{-3.5}\text{ cm}^2/\text{g}$ and there is no scattering ($\sigma_s=0$).

The upstream conditions for the gas are set to a density of $\rho=1\text{g/cc}$, $v=0$ and $T_m=T_r=0.122\text{keV}$. The downstream conditions are determined from hydrodynamic jump conditions solution~\citep{shock} to be a density of $\rho=2.29\text{g/cc}$, $v=-1.95\cdot 10^7\text{cm/s}$ and $T_m=T_r=0.253\text{keV}$. The initial profile of the gas is set to the steady state semi-analytical solution, using a \textsc{matlab} script provided by~\cite{kolin} that implements the solution strategy from ~\cite{shock}. The system is evolved for $5$ns, in which the shock has enough time to propagate several times its size. The domain $x\in[-0.21,\; 0.7]$cm is divided into 1024 equally spaced cells and the boundary conditions are set to the upstream/downstream values. The IMC and DIMC runs create $\approx 10^6$ new photons each time step and limit the total number of photons to be $10^7$. For the ISMC runs we limit the total number of particles to be $10^8$ in order to get good statistics and even then, the large ratio between the heat capacity of the gas and the radiation prevents us from having sufficient statistics in the radiation field. In the DIMC runs we use 20 material particles per cell.
\begin{figure}
\centering
(a)\includegraphics*[width=7.8cm]{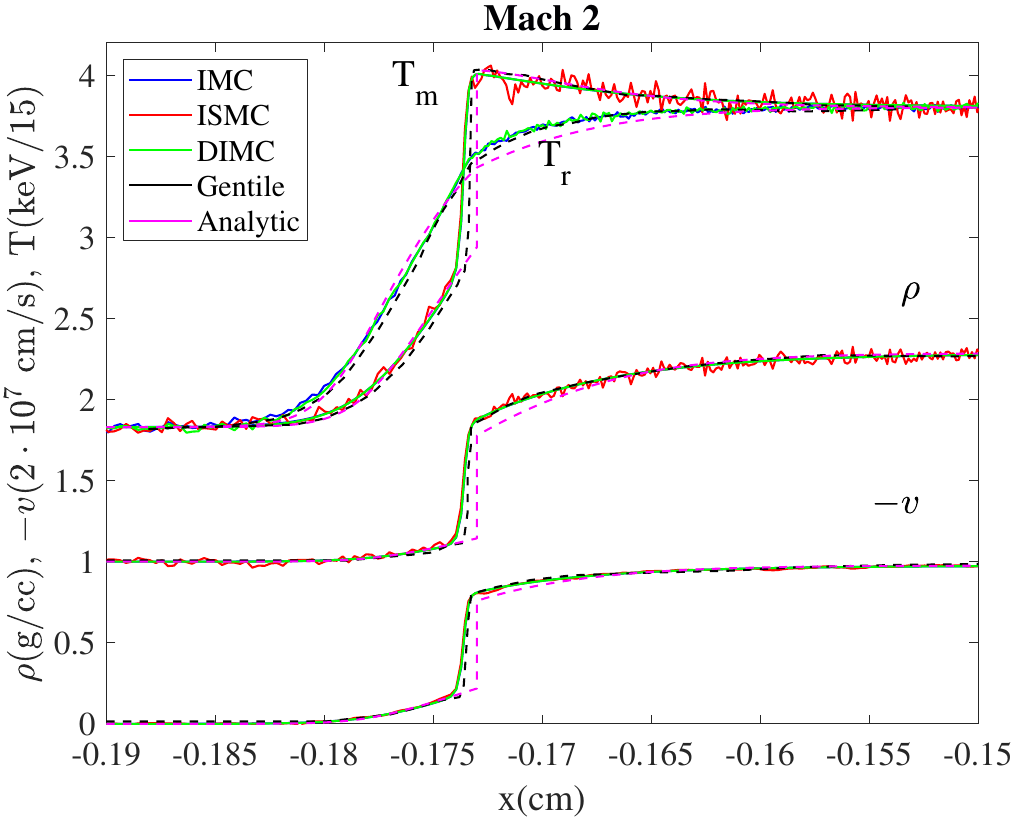}
(b)\includegraphics*[width=8cm]{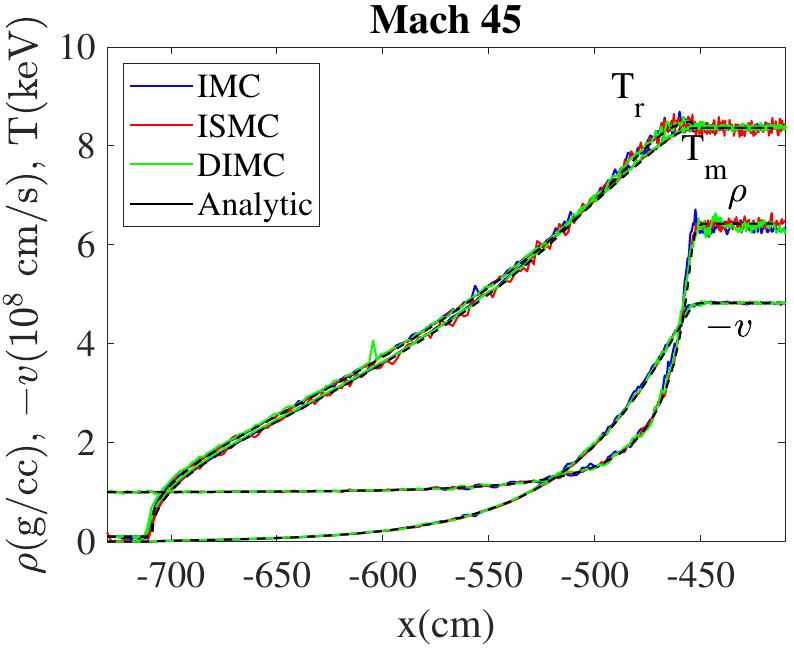}
\caption{(a) The hydrodynamical profiles (material and radiation temperatures, density and velocity) of $\M=2$ radiative shock problem. (b) Same for the high Mach problem, $\M=45$, where the physics is close to LTE. The solutions are compared to the semi-analytic solution of~\cite{shock} using the \textsc{matlab} script provided by~\cite{kolin}. In addition, in the $\M=2$ problem, a transport solution ($S_N$, taken from~\cite{gentile2019}) which differs from the diffusion solution in this non-LTE problem is presented for comparison.}
\label{fig:Mach2}
\end{figure}

Fig.~\ref{fig:Mach2}(a) shows the hydrodynamical properties for the different methods compared to the analytical result (the magenta curves). In this problem, there is a deviation between the exact transport and the 2-temperature diffusion semi-analytic solution, so we add the $S_N$ solution~\cite{Sn2,gentile2019} (black curves, which is in very good agreement also with Gentile's IMC results), to benchmark our Monte-Carlo results. %{\bf In principle, \cite{Sn2} have calculated an exact semi-analytic solution for this problem, but unfortunately we do not have access to the source code of the solution.}
Overall there is a good agreement between the three different methods with the $S_N$ results. The deviation between the diffusion solution and all the Boltzmann solutions are larger than the deviation between the Boltzmann solutions themselves.
The noise level of IMC and DIMC is comparable, while in the ISMC method the noise is significantly higher.

\subsection{Mach 45 ($\M=45$) Case}
For our next test we run the Mach 45 radiating shock presented in~\cite{nick_mmc,gentile2019}.
The gas is assumed to be an ideal gas with an adiabatic index of $\gamma=5/3$, specific heat capacity of $1.25\cdot 10^8\textrm{ erg/g}\cdot\text{K}$, the scattering opacity is $\sigma_s=0.4006\rho\text{ cm}^2/\text{g}$ and the absorption opacity is $\sigma_a=0.0142\rho^2 (T/\text{keV})^{-3.5}\text{ cm}^5/\text{g}^2$. The upstream conditions for the gas are a density of $\rho=1\text{g/cc}$, $v=0$ and $T_m=T_r=0.1\text{keV}$. The downstream conditions are determined from the hydrodynamical jump conditions solution~\citep{shock} to be a density of $\rho=6.43\text{g/cc}$, $v=-4.82\cdot 10^8\text{cm/s}$ and $T_m=T_r=8.36\text{keV}$. As in the $\M=2$ benchmark, the initial profile of the gas is set to the steady state solution using the \textsc{matlab} script provided by~\cite{kolin} reproducing~\cite{shock} solutions, and the system is evolved for $8\cdot 10^{-7}$s, in which the shock has enough time to propagate several times its size. The domain $x\in[-879,\; 91]$ is divided into 512 equally spaced cells and the boundary conditions are set to the upstream/downstream values. The IMC and DIMC runs create $\approx 4\cdot 10^5$ new photons each time step and limit the total number of photons to be $5\cdot 10^6$. For the ISMC runs we limit the total number of particles to be $5\cdot 10^7$ in order to get sufficient statistics. In the DIMC runs we use 30 material particles per cell.

Fig.~\ref{fig:Mach2}(b) shows the hydrodynamical properties for the different methods compared to the analytical result. Overall there is a good agreement between the three different methods, the noise level of IMC and DIMC is comparable, while in the ISMC method the noise is significantly higher. The agreement between our Monte-Carlo schemes to the radiative shocks benchmarks shows that the DIMC runs perfectly also in radiative hydrodynamic including MMC corrections. However, these problems are not opaque enough to show the strength of DIMC compared to IMC and ISMC. In Sec. ~\ref{45op} we rerun the $\M=45$ case, only with  ahigher opacity, showing the difference between the Monte-Carlo schemes.

\subsection{Mach 45 ($\M=45$) High Opacity Case}
\label{45op}
\begin{figure}
\centering
\includegraphics*[width=8cm]{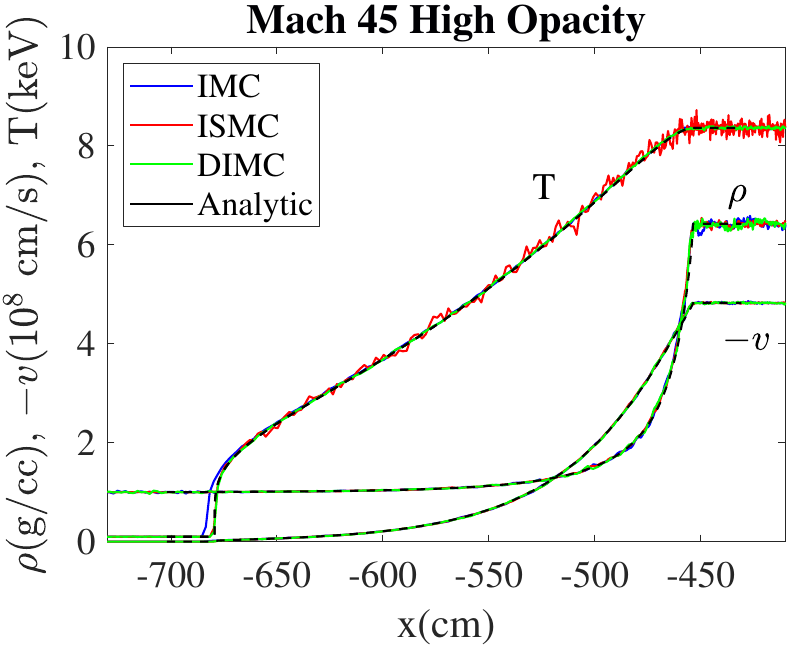}
\caption{The hydrodynamical profiles (material temperature, density and velocity) of $\M=45$ in highly-opaque problem radiative shock problem using 512 spatial cells. The solutions are compared to the semi-analytic solution of~\cite{shock}.}
\label{fig:Mach45Opac}
\end{figure}

We now repeat the previous $\M=45$ benchmark while increasing the absorption opacity by a factor of 100. 
We benchmark our solution against the semi-analytical reference solution~\citep{kolin,shock}. 
First, Fig.~\ref{fig:Mach45Opac} shows the hydrodynamical properties for the different methods using 512 (spatially converged) cells compared to the analytical solution.All the Monte-Carlo schemes yield good results with the semi-analytical solution.

\begin{figure}
\centering
(a)\includegraphics*[width=5.5cm]{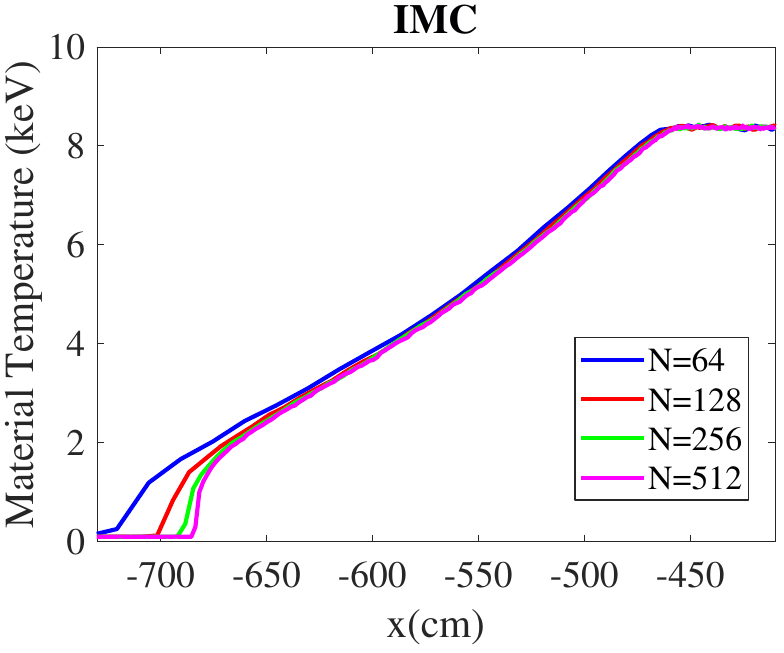}
(b)\includegraphics*[width=5.5cm]{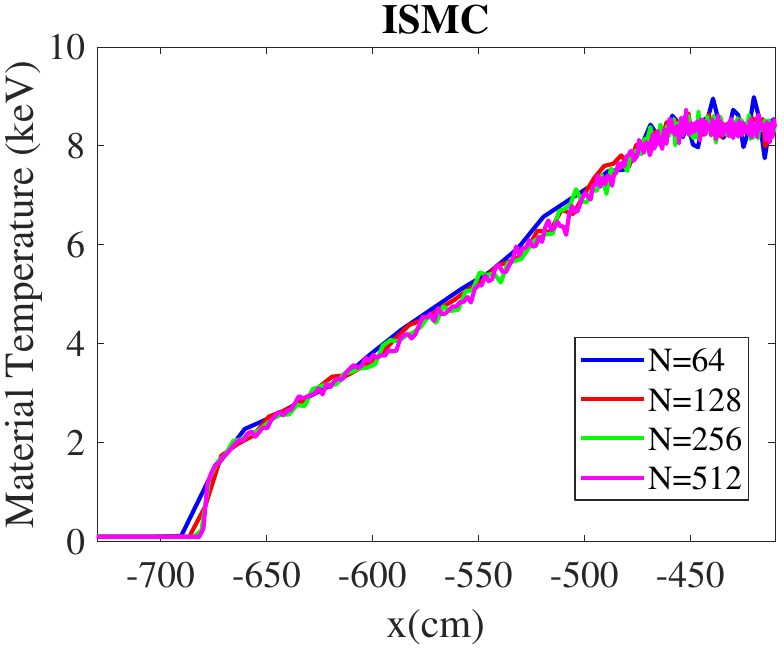}
(c)\includegraphics*[width=5.5cm]{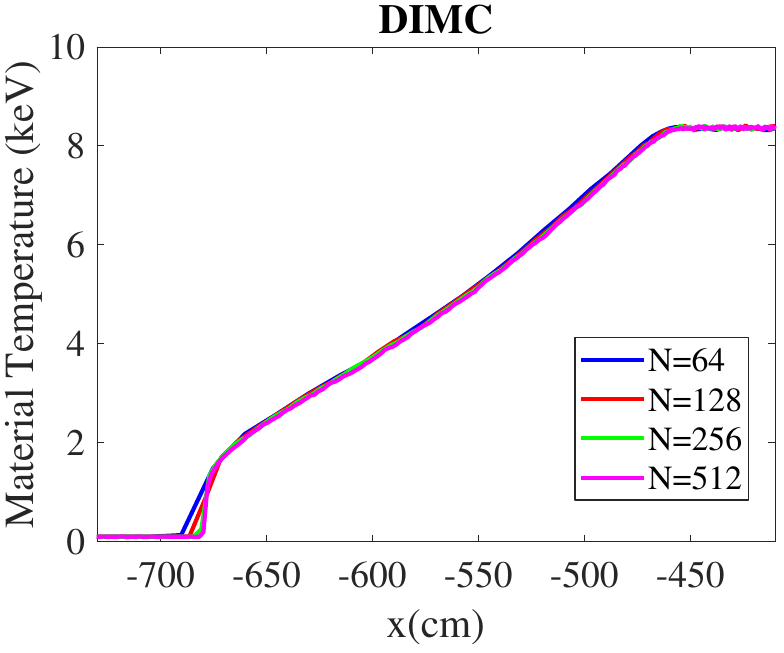}
\caption{The high-opacity $\M=45$ problem gas temperature for different spatial resolutions for (a) IMC, (b) ISMC and (c) DIMC.}
\label{fig:Mach45OpacIMC}
\end{figure}

Next, we run additional runs varying the spatial resolution while keeping the number of photons/particles per cell fixed. Fig.~\ref{fig:Mach45OpacIMC} shows the material temperature for different spatial resolutions using the IMC (a), ISMC (b) and DIMC (c) methods. Using classic IMC (Fig.~\ref{fig:Mach45OpacIMC}(a)), as the resolution decreases, the teleportation error increases and the heat wave travels too fast into the upstream region. Using ISMC (Fig.~\ref{fig:Mach45OpacIMC}(b)) and DIMC (Fig.~\ref{fig:Mach45OpacIMC}(c)), there are no teleportation errors even when the spatial resolution is very low. Again, we note that comparing ISMC versus DIMC, the ISMC implementation suffers from a higher noise level than the DIMC run.

\section{Conclusions}
Monte-Carlo algorithms for modeling radiative transfer problems in multi-dimensional problems are increasing in popularity, as computational capabilities grow. This is due to the high accuracy of this methods (when sufficient statistics is available), compared to low-order approximations such as various diffusion or Eddington factor approximations methods. The classic MC implementation of radiative transfer is IMC~\citep{IMC}, which suffers from teleportation errors in optically-thick regions, using finite spatial and temporal resolutions. The recently introduced implicit algorithm, ISMC~\citep{ISMC} yield no teleportation errors, however, it gives rise to noisier results due to its discrete treatment in transferring the energy between radiation and material.
In this study, we have presented a new method to solve the radiation transfer equations. The new method, Discrete IMC (DIMC), is the offspring of both IMC and ISMC. From IMC we adopt its relatively smooth material temperature properties and continuous energy deposition. From ISMC we adopt the discrete nature of material particles, that is crucial in preventing teleportation in optically thick regions. DIMC inherits all of the good qualities from its parents while side stepping their major drawbacks; teleportation in IMC and increased noise in ISMC. As opposed to ISMC, the decoupling of the energy scales in DIMC between the radiation photons and material particles, allows quick convergence in problems where there is a large ratio between the heat capacities of the gas and the radiation.

The new DIMC algorithm was tested against several benchmarks, both one-dimensional and two-dimensional problems that include complex radiation flows. In addition, it was tested against radiative hydrodynamics benchmarks, in different optical depths.
In all of our presented radiative transfer benchmarks the performance of DIMC either superseded that of IMC and ISMC, or was comparable. We have found that the DIMC, that uses the same rationale of material particles as ISMC, yields no teleportation errors in all the examined benchmarks, with or without hydrodynamics. However, the noise level of DIMC is better than ISMC for a given statistic, and yields smoother results, like IMC. The added run time from having additional material particles is not large, and at worse, DIMC runs less than a factor of two slower than IMC, and in several cases even faster.

The new DIMC method can be easily extended to non-gray cases, in the same manner as traditional IMC is extended. Additionally, random walk methods that accelerate photon propagation in optically thick media~\citep{RW}, are fully compatible with DIMC and can be implemented with no changes. Concerning a 3D possible extension, there is a question of what is the number of material particles that yield a desired accuracy of the solution. We estimate that about $\approx100$ material particles per cell should give good results in 3D. The computational cost of the added material particles should not be high since the CPU cost is determines mostly on how many {\em{photons}} there is per cell, rather than material particles.

\section*{Acknowledgement}
We would like to thank Assaf Kolin for permission to use his semi-analytic solver code for the Loweri's two-temperature radiative shock problem. We would also like to thank Gaël Poëtte and Nicholas A. Gentile for taking the time to read and comment on this paper, as well as fruitful discussions.
%Gaël Poëtte for taking the time to read and comment on this paper, as well as fruitful discussions.

\bibliography{sample631}{}
\bibliographystyle{aasjournal}

%% This command is needed to show the entire author+affiliation list when
%% the collaboration and author truncation commands are used.  It has to
%% go at the end of the manuscript.
%\allauthors

%% Include this line if you are using the \added, \replaced, \deleted
%% commands to see a summary list of all changes at the end of the article.
%\listofchanges

\end{document}